\documentclass[fleqn,twoside]{article}
\usepackage{espcrc2}
\usepackage{graphicx} 
\newcommand{\AmS}{{\protect\the\textfont2
  A\kern-.1667em\lower.5ex\hbox{M}\kern-.125emS}}
\input babarsym

\title{Recent results on semileptonic decays at \babar\ }

\author{J. Serrano\address{Laboratoire de l'acc\'el\'erateur Lin\'eaire, \\
       B\^atiment 208, BP34,  91898 Orsay Cedex, France}, representing the \babar\ Collaboration}

\begin{document}

\begin{abstract}
Some recent \babar\ results on semileptonic decays are presented.
They focus on the determination of the CKM matrix elements $|V_{ub}|$ 
and $|V_{cb}|$ in inclusive and exclusive $b \to u \ell \nu$ and  $b \to c \ell \nu$ 
decays, and on form factors measurement in exclusive $c \to s \ell \nu$ decays. 
\end{abstract}

\maketitle

\section{Introduction}
Semileptonic decays play a crucial role in the determination of the unitarity triangle parameters: decays of the $b$ quark give access to the CKM matrix elements $|V_{ub}|$ and
 $|V_{cb}|$, while charm  decays provide a way to validate lattice QCD  computations through form factors measurements. Such calculations provide 
theoretical inputs that are used, especially,  in the b sector.
A lot of new results have been obtained by the \babar\ collaboration during the last years,
thanks to the large $b\bar{b}$ and $c\bar{c}$ production cross-sections and to the large
recorded statistics. Some of these measurements are presented here.

\section{$\bar{B} \to X_c \ell^- \bar \nu_\ell$ decays}

\subsection{Inclusive analysis: measurement of moments of the hadronic-mass and of the lepton energy spectrum}
In the context of Heavy Quark Expansion (HQE),
measurements of moments of the  hadronic-mass and of the lepton-energy spectra in inclusive 
$\bar{B} \to X_c \ell^- \bar\nu_\ell$, and measurements of moments of the photon-energy spectrum
in $\bar{B} \to X_s \gamma$ decays, are used to determine precisely $|V_{cb}|$, the quark
masses $m_b$ and $m_c$ and the heavy-quark parameters. In the analysis presented here \cite{ref:c_incl},
$232$ millions of $B\bar{B}$ pairs are used to obtain a new measurement of  hadronic 
mass moments $\langle m_X^k\rangle$ with $k=1,...,6$  as well as a first determination
of  mixed hadron mass-energy moments $\langle n_X^k\rangle$ with $k=2,4,6$. All
moments are given for different cuts on the minimum momentum of the charged lepton, varying between $0.8\gevcc$ and  $1.9\gevcc$, in the rest frame of the $B$ meson.
Events with one $B$ meson fully reconstructed in a hadronic decay are used, the semileptonic decay of the second $B$ meson  is identified by the presence of an electron or a muon.
The hadronic system $X_c$ is reconstructed from the remaining particles in the event and the hadronic mass is calculated from the reconstructed four-momenta as $m_X=\surd{p_{X_c}^2}$.
To extract unbiased $\langle m_X^k\rangle$ moments, correction functions defined from the simulation are used. They relate
moments of the measured mass and moments of the true underlying mass and
depend on the resolution and total multiplicity of the hadronic system $X_c$. The same
procedure is used to extract the mixed   moments $\langle n_X^k\rangle$.
The measured hadronic mass moments shown in \cite{ref:c_incl} agree with previous 
measurements and present significantly smaller statistical uncertainties, which are 
smaller than the systematic uncertainties, than the previous \babar\ measurements.
A combined fit in the kinetic scheme to  hadronic mass moments, 
to measured moments of the lepton-energy spectrum \cite{ref:c_incl1}
and to moments of the photon energy spectrum in $\bar{B} \to X_s \gamma$ \cite{ref:c_incl2,ref:c_incl3} ,
yields preliminary results for $|V_{cb}|$,  $m_b$, $m_c$, the total semileptonic
branching fraction ${\cal B}(\bar{B} \to X_c \ell^- \bar\nu_\ell)$ and perturbative HQE parameters
in agreement with previous determinations. In particular, the following values are found:
 $|V_{cb}|=(41.88\pm0.81)\cdot 10^{-3}$ and $m_b=(4.552\pm0.055)\gevcc$.

\subsection{Exclusive analysis: $B^0\to D^{*-}\ell^+ \nu_\ell$}
\begin{figure}[!htbp]
\begin{center}
\includegraphics[width=5cm]{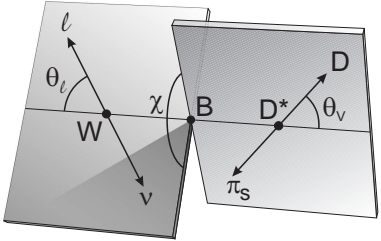}
\caption{Definition of the kinematic variables for the $B^0\to D^{*-}\ell^+\nu_\ell$ decay.}
\label{fig:BtoDstarvar}
\end{center}
\end{figure}

The study of the  $B^0\to D^{*-}\ell^+\nu_\ell$ decay  allows the
simultaneous determination of $|V_{cb}|$ and of the form factors parameters characterizing 
the effects of strong interaction in this decay. There are two axial form factors, $A_1$ and $A_2$, and one vector form factor, $V$, which
depend on $q^2$, the mass squared of the $\ell^+\nu_\ell$ system. The heavy quark effective
theory (HQET) predicts that these form factors are related to each other through
heavy quark symmetry (HQS), but HQET leaves three free parameters, which must be extracted
from experiment.
This decay depends on four kinematic variables: $w$, which is related to   $q^2$  by $w=\frac{m_B^2+m_{D^*}^2-q^2}{2m_B m_{D^*}}$; and the three decay angles ($\theta_\ell$, $\theta_V$,  $\chi$) defined in Fig. 
\ref{fig:BtoDstarvar}.

The Lorentz structure of the $B^0\to D^{*-}\ell^+\nu_\ell$ decay amplitude can be expressed in
terms of three helicity amplitudes which correspond to the three polarization states
of the $D^*$. For low-mass leptons, these amplitudes are expressed in terms of the three
functions $h_{A_1}(w)$, $R_1(w)$ and $R_2(w)$, related to the form factors $A_1$, $A_2$ and $V$.
The analysis reported here \cite{ref:c_excl},  uses the following expressions for the form factor
parameterization \cite{ref:c_excl1}:
\begin{eqnarray}
\label{eq:Cap}
h_{A_{1}}(w) & = &
h_{A_{1}}(1)
\big[ 1-8\rho^{2} z+(53\rho^{2}-15)z^{2} \nonumber \\ 
& &  -(231\rho^{2}-91)z^{3}\big], \nonumber \\ 
R_{1}(w) & = & R_{1}(1)-0.12(w-1)+0.05(w-1)^{2}, \nonumber \\ 
R_{2}(w) & = & R_{2}(1)+0.11(w-1)-0.06(w-1)^{2}, \nonumber 
\end{eqnarray}
\noindent
where  $z=[\sqrt{w+1}-\sqrt{2}]/[\sqrt{w+1}+\sqrt{2}]$.
The three parameters $\rho^{2}$, $R_{1}(1)$, and
$R_{2}(1)$, cannot be calculated; they must be extracted from data.
In \babar\, this analysis is performed using a sample of
$79~fb^{-1}$. Events that contain a $D^{*-}$ candidate and an oppositely charged electron
or muon with momentum in the range $1.2<p_{\ell}<2.4~\gevc$ are selected. The
$D^{*-}$ is reconstructed in the decay channel  $\Dstarm \ra \Dzb \pi^-$, 
with the \Dzb\ decaying to $\Kp \pim,~\Kp \pim \pip\pim$, or $\Kp \pim\piz$. About 
$52,800$ $B^0\to D^{*-}\ell\nu$ decays are reconstructed. 
The value of $\mathcal{F}(1)|V_{cb}|$ and of the three form factors parameters are extracted using
a combined fit of three one-dimensional ($\chi$ is practically insensitive to the form factors parameters)  binned distributions with a bin-by-bin background
subtraction. 
%The three kinematic variables chosen are $w$, $\cos\theta_\ell$ and   $\cos\theta_V$ as
The background is estimated from data
independently for each variable.  
Combining  results from this analysis with  the ones contained in the previous \babar\ publication 
\cite{ref:c_excl2} and taking into account the correlation between them,
the following values are obtained:
\begin{eqnarray*}
\mathcal{F}(1)|V_{cb}| & = & ( 34.4 \pm 0.3 \pm 1.1) \times 10^{-3}  \\
\rho^2 & = & 1.191 \pm 0.048 \pm 0.028 \\
R_1 & = & 1.429 \pm 0.061 \pm 0.044 \\
R_2 & = & 0.827 \pm 0.037 \pm 0.022.
\end{eqnarray*}
The corresponding branching fraction is $\mathcal{B}(B^0\to D^{*-}\ell^+\nu_\ell) = (4.69 \pm 0.04 \pm 0.34 )\%$.
These results supersede all previous \babar\ measurements of the form factors parameters, of the exclusive branching fraction and of $|V_{cb}|$ extracted from this decay.

\section{$\bar{B} \to X_u \ell^- \bar\nu_\ell$ decays}
\subsection{Inclusive analysis}
The analysis of inclusive $\bar{B} \to X_u \ell^- \bar\nu_\ell$ decays allows the determination of the
CKM matrix element $|V_{ub}|$ through the measurement of the decay rate.
The experimental challenge for this analysis is to separate the signal from the
50 times larger  $\bar{B} \to X_c \ell^- \bar\nu_\ell$ decays. Thanks to the mass difference between the
$u$ and $c$ quark, several  regions of phase space can be defined where this background is suppressed.
The measured partial branching fractions, $\Delta \BR(\bar{B} \to X_u \ell^- \bar\nu_\ell)$, in these selected regions can then be related to
 $|V_{ub}|$ thanks to QCD calculations in  the Operator Product Expansion (OPE) framework.
This analysis has been done by the \babar\ collaboration using $347.4~fb^{-1}$ \cite{ref:u_incl}. Events with one 
of the $B$ meson fully reconstructed in a hadronic decay are selected. The semileptonic decay of the 
second $B$ meson is identified by the presence of an electron or a muon with momentum in the center of
mass frame greater than 1 $\gevc$. Three kinematic variables are used to select three 
different  regions of phase space: $M_X$, the invariant mass of the hadronic system,
$q^2$  and $P_+\equiv E_X-|\vec{P}_X|$
where  $E_X$ and $\vec{P}_X$ are the energy and momentum of the hadronic system in the $B$ rest frame.
The distribution of these variables are extracted performing fits to the $m_{ES}$ \footnote{$m_{ES}=\sqrt{s/4-\vec{p}^2_B}$, where $\sqrt{s}$ is the total energy in the
$\Upsilon(4S)$ center-of-mass frame and $\vec{p}_B$, the momentum of the $B$ candidate in the same frame.} distribution of
the reconstructed $B$, for subsamples of events in individual bins for each of the kinematic variables.
The partial branching ratios are measured for $M_X<1.55\gevcc$, $P_+<0.66\gevc$ and $(M_X<1.7\gevcc,q^2>8
(\gevcc)^2)$. Actually, in order to reduce  systematic uncertainties,  ratios of partial 
branching fractions to the total semileptonic branching fraction are measured.
Results of the fitted number of events, $\Delta \BR(\bar{B} \to X_u \ell^- \bar\nu_\ell)$ and the
corresponding values of $|V_{ub}|$ for the three kinematic regions 
can be found in \cite{ref:u_incl}.
The partial branching ratios are translated to $|V_{ub}|$ using recent QCD calculations.
The analysis which uses the $M_X$ variable leads to a very accurate determination of $|V_{ub}|$,
with a $9\%$ total uncertainty.

\subsection{Exclusive analysis: $B^0\to \pi^{-}\ell^+\nu_\ell$}
The rate of the exculsive $B^0\to \pi^{-}\ell^+\nu_\ell$ is proportional to $|V_{ub}f_+(q^2)|^2$,
where the form factor $f_+(q^2)$ depends on the momentum transfered squared $q^2$.
Several theoretical calculations, as light cone sum rules or lattice QCD  provide values of this form factors for different
$q^2$ range, which allows the measurement of $|V_{ub}|$ from experimental data. 
Uncertainties on these calculations still dominate the errors on the computed
values of $|V_{ub}|$, but, if a large statistics  is available, the
 data can  be used to discriminate between the various calculations
by precisely measuring the  $f_+(q^2)$ shape.
The \babar\ collaboration recently published an analysis of the $B^0\to \pi^{-}\ell^+\nu_\ell$ 
decay \cite{ref:u_excl} using  an original method based on a loose neutrino reconstruction technique. 
Using $206~fb^{-1}$, the $B$ meson candidate is reconstructed using $\pi^{\pm}$ and $\ell^{\mp}$
together with the event's missing momentum as an approximation to the signal neutrino momentum.
The decay of the second $B$ meson is not explicitely reconstructed. This leads to a large
 signal efficiency while having a good $q^2$ resolution. A total of  $5072\pm 251$ signal 
events are obtained and the partial branching fractions $\Delta \BR(B^0\to \pi^{-}\ell^+\nu_\ell,q^2)$
are measured in 12 $q^2$ bins. The  $\Delta  \BR$ distribution
together with theoretical predictions  is shown in \cite{ref:u_excl}. Using \cite{ref:u_excl1} in the range $q^2>16\gev^2$,
 the value of $|V_{ub}|$ is obtained: $|V_{ub}|=3.4\pm0.2\pm0.2^{+0.6}_{-0.4}$, where the last uncertainty is due to the normalization of the form factor. A  precise determination
of the total branching fraction is also obtained: $\BR(B^0\to \pi^{-}\ell^+\nu)=(1.46\pm0.07\pm0.08)\times 10^{-4}$.

\section{$c \to s \ell^+ \nu_\ell$ decays}
The \babar\ collaboration has obtained precise measurements of the form factors parameters for two charm 
semileptonic decays: $D^0\to K^-e^+ \nu_e$ \cite{ref:kenu} and $D_s^+ \to K^+K^- e^+\nu_e$ \cite{ref:phienu}. 
The CKM matrix element involved in these decays,  $V_{cs}$ is known precisely if we assume the  unitarity of the CKM matrix.
The analysis technique is similar for both channels, the charm decay is reconstructed
in $e^+e^-\to c\bar{c}$ events, the second charm meson is not explicitely reconstructed.
The difference of shape between $c\bar{c}$ and $b\bar{b}$ is used to discriminate 
signal from $b\bar{b}$ background events.

\subsection{$D^0\to K^-e^+ \nu_e$ decay}
The rate of this exclusive decay is proportional to $|f_+(q^2)|^2$. The models used to parameterized the $q^2$ dependence of the form factor are the BK ansatz \cite{ref:BK}, where the parameter to be fitted is $\alpha$ and the simple pole model, for which the mass of the pole is fitted.
This analysis has been done with $75~ fb^{-1}$. $D^0$ from $D^{*+}\to D^0\pi^+$ decays are used,
and a sample of about 74,000 signal events is selected. The true $q^2$ distribution is 
obtained using an unfolding algorithm.
It is then fitted with the different models. The determined pole mass is $m_{pole}=1.884\pm0.012\pm0.015\gevcc$, which is lower than the expected value ($m_{pole}=m_{D_s^*}=2.112\gevcc$), excluding the simple pole model.
The modified pole mass parameter $\alpha=0.38\pm0.02\pm 0.03$. This value is lower than 
the lattice QCD determination \cite{ref:kenulatt}($\alpha=0.50\pm0.04$).
In order to obtain the absolute normalization of the form factor, the $D^0\to K^- e^+ \nu_e$
branching fraction is measured relative to the reference decay channel $D^0\to K^-\pi^+$. 
The extracted value of $f_+(0)$ is found to be $f_+(0)=0.727\pm 0.005 \pm 0.007 \pm 0.005$, 
where the uncertainties are statistical, systematic and from external inputs, repsectively.
This value is in agreement with the lattice result  \cite{ref:kenulatt} ($f_+(0)=0.73\pm 0.03 \pm0.07$).

\subsection{$D_s^+ \to K^+K^- e^+\nu_e$}
As for the  $B^0\to D^{*-}\ell^+\nu_e$, this decay depends on four variables ($q^2$ and three decay angles) and on three form factors, $A_1$, $A_2$ and $V$, for which we assume a $q^2$ dependence
dominated by a single pole:
\begin{eqnarray*}
V(q^2)=\frac{V(0)}{1-q^2/m_V^2};~A_{1,2}(q^2)=\frac{A_{1,2}(0)}{1-q^2/m_{A}^2}. \\
\end{eqnarray*}
Events with a $K^+K^-$ mass in the range $1.01-1.03\gevcc$ are selected, and except for a small
S-wave contribution, they correspond to $\phi$ meson.
Using $214 fb^{-1}$ of data, the number of selected signal events  is 25341, which greatly exceeds any previous measurement and allows the determination of the pole mass $m_A$ in addition to the usual form factors  parameters $r_2=A_2(0)/A_1(0)$, $r_V=V(0)/A_1(0)$. These parameters
are extracted using a binned maximum likelihood fit to the four-dimensional decay distribution.
The sensitivity to $m_V$ is weak and this parameter is fixed to $2.1\gevcc$. The following 
values are obtained: $r_2=0.763\pm0.071\pm0.065$, $r_V=1.849\pm0.060\pm0.095$, $m_A=2.28^{+0.23}_{-0.18}\pm0.18\gevcc$.  
\babar\ also finds a first evidence for a small S-wave contribution associated 
with $f_0 \to K^+K^-$ decays corresponding to $(0.22^{+0.12}_{-0.08}\pm0.03)\%$ of the
$K^+K^- e^+\nu$ decay rate. 
Measuring the $\Ds \rightarrow K^+ K^- e^+ \nu_e$ branching fraction
relative to the decay $\Ds \rightarrow  K^+ K^- \pi^+$, the absolute normalization
is obtained and $A_1(0) = 0.607 \pm 0.011 \pm 0.019 \pm 0.018$.
Lattice  calculations for this channel have
been done in the quenched approximation. They agree with the present experimental result
of  $A_1(0)$, $r_2$ and $m_A$, but are lower than the measured value 
of $r_V$.

\section{Conclusion}
This review of recent \babar\ measurements, altough non-exhaustive, 
underlines the great effort ongoing at $B$-factories in the understanding of
semileptonic decays. One can remark that
the dominant uncertainties on the determination of the CKM parameters are  
often coming from theoretical inputs.

\end{document}